# Strategies in object-oriented design[1]

Sophie Chatel and Françoise Détienne

*INRIA, Project of Ergonomics Psychology, Rocquencourt, France*

This paper presents a study aiming to analyse the design strategies of experts in object-oriented programming. We report an experiment conducted with four experts. Each subject solved three problems. Our results show that three strategies may be used in program design according to the solution structure. An object-centred strategy and a function-centred strategy are used when the solution has a hierarchical structure with vertical communication between objects. In this case, the plan which guides the design activity is declarative. A procedure-centred strategy is used when the solution has a flat structure with horizontal communication between objects. In this case, the plan which guides the design activity is procedural. These results are discussed in relation with results on design strategies in procedural design. Furthermore, our results provide insight into the knowledge structures of experts in object-oriented design. To conclude, we point out limitations of this study and discuss implications of our results for Human-Computer Interaction systems, in particular for systems assisting experts in their design activity.

PsychINFO classification: 2340 Cognitive Processes

The psychology of problem-solving has almost always been focused on well-structured problems. Software design problems have been described as 'ill-structured' or 'ill-defined' problems (Guindon 1990; Pennington and Grabowski 1990; Visser and Hoc 1990). Important features of design problems are:
— incomplete and ambiguous specification of goals;
— no predetermined solution path;

---

[1] This research was partly supported by the SCALE Esprit III project 6334 (System Composition and Large Grain Component Reuse Support).
*Correspondence to :* S. Chatel and F. Détienne, Project of Ergonomics Psychology, INRIA, Domaine de Voluceau, Rocquencourt - BP 105, 78153, Le Chesnay, France.

—   the need to integrate multiple knowledge domains. Specification and constraints come from various knowledge domains and have to be translated into a specific knowledge domain, the programming domain;
—   constraints are often conflicting;
—   there is no definite criterion for testing any proposed solution and various solutions are acceptable, one being possibly more satisfactory in one dimension, another in another dimension.

As remarked by Davies (1993), much of the literature concerned with understanding the nature of programming skill has focused explicitly upon the declarative aspects of programmer's knowledge. One major limitation of many of these studies is that they fail to consider the way in which such knowledge is used or applied in programming tasks like the design task.

This paper presents a study aiming to analyse a main aspect of expertise in object-oriented design (OOD): the design strategies used by experts.

**Shifting paradigm**

Many empirical studies have been conducted to analyse the design activity and the knowledge possessed by experts in programming. Most of these studies have been conducted with the procedural paradigm (see for example, Adelson and Soloway 1985; Détienne 1990a, b; Guindon 1990; Rist 1986; Robertson and Yu 1990; Soloway et al. 1982; Soloway and Ehrlich 1984).

Lately, researchers have been interested in the design activity with a more recent paradigm, the OO paradigm. Some recent empirical studies on OOD (see for example, Détienne 1995; Kim and Lerch 1992; Lange and Moher 1989; Pennington et al. 1995; Rist to appear) reflect this interest.

There is an important difference between the procedural paradigm and the OO paradigm. In procedural programs, data and procedures are separated whereas, in OO programs, they are integrated. Objects are program entities which integrate a structure defined by a type as well as functionalities. Objects are instances of classes. A class describes the attributes (called instance variables in Smalltalk) and operations (i.e. methods) common to a set of objects. A class is defined as a type of abstract data. A method is a function attached to a class that describes a piece of the behaviour of the objects which are instances of this class. Classes may be organised according to a hierarchical structure; classes inherit the properties (static and functional) of their superclass(es).

Our study involves experts with an OO language, Smalltalk. The focus will be on the characteristics of the design strategies used by experts and on the conditions which influence the choice of one strategy rather than another. Previous studies on OOD have been focused mostly on the reuse activity (Burkhardt and Détienne 1995; Détienne 1991; Lange and Moher 1989; Lewis et al. 1991; Rosson and Carroll 1993), on the comparison between the two paradigms as regards the organisation of the design activity (Pennington et al. 1995), and on the expert-novice comparison, in particular the transfer effects when switching from procedural to OOP (Chatel et al. 1992; Détienne 1990a; Détienne 1995; Pennington et al. 1995). To date, there have been few works focused on the characteristics of design strategies with the OO paradigm (Détienne 1995; Rist to appear).

**Design strategies**

Most studies on software design have identified the same global control strategy: decomposition of the problem into subproblems. In procedural design, the distinction between different types of problem decomposition has been highlighted. Ratcliff and Siddiqi (1985) distinguished between data-driven strategies and goal-driven strategies which lead to different problem decompositions. Hoc (1983) proposed a more complex framework to classify problems and strategies. Two dimensions are distinguished: procedural versus declarative, and prospective versus retrospective. In procedural problems, the program structure is strongly constrained by the procedure structure; this structure guides the solution development. In declarative problems, the program structure is strongly constrained by the data structure; this structure guides the solution development. Furthermore, the solution may be developed in a prospective (or forward) way or in a retrospective (or backward) way.

These distinctions have been made in the context of procedural design. A question is to analyse whether or not the same strategies are involved in OOD and the conditions which influence the use of one strategy rather than another. In our study, we will focus our analysis on the distinction between plan-based strategies. Two main plans may guide the design activity. Either the plan is declarative, where static characteristics like objects, main goals, and typical functions guide the solution development, or the plan is procedural, where dynamic characteristics of the procedure guide the solution development.

In previous work (Détienne 1995), we showed that expert designers in OOD tend to decompose their solutions around the objects. The strategy was declarative. It was found that methods names were generated mainly according to their being members of the same class whatever the problem type (declarative versus procedural). This result questioned the classification of problems made for procedural design and its relevance for OOD.

In the present study, we will analyse whether or not another dimension of the problem may influence the choice of design strategies in OOD. It is likely that this new dimension may characterize not only the data structure (or objects), on one side, or the procedure, on the other side, but rather the way they are associated. The structure of the solution and the way objects communicate within this structure is an important feature of the solutions in OOD. We could assume that various design strategies may be used for problems which differ according to the static structure of their solution, i.e. hierarchical or flat, and according to the way objects communicate within this structure. Of course, it could be argued that the OO paradigm encourages the development of hierarchical solution structures rather than flat solution structures but our point is that for any large software development, we could distinguish parts of the solution with a hierarchical structure from parts with a relatively flat structure. In this case, it is likely that various design strategies are used for developing parts which have different structures.

In order to analyse the design strategies as related to the solution structure, we will use three problems which have been shown to differ according to the structure of their solution. In a pre-experiment, an expert who teaches OO programming at a university had produced solutions to these problems. For two of them the solution was hierarchical whereas for the third problem the solution was flat.

Furthermore these solutions differed according to the way objects communicated with each other.

The plan which guides the activity will be assessed in this study by the order of generating code. The generation order should provide insight into the structure of the plans which guide the design activity. We will analyse whether the order in which lines of code were generated is guided by static characteristics like objects or functions or by a dynamic structure, i.e., the calling method structure which determines the order of execution of the procedure.

In the next section we report one experiment on OO software design. Results of this experiment are discussed and the implications of these results for HCI systems, in particular for systems assisting experts in their design activity, are developed in the concluding sections.

**Methodology**

*Subjects*

Four experts in Smalltalk participated in this experiment. They had from 3 to 5 years of experience in Smalltalk. All the subjects were familiar with 4 to 6 programming languages (at least one procedural language, one functional language and one object-oriented language).

*Material*

Three problems were used: a management problem and two isomorphic problems[2]. The management problem calculates the number of standard model cars which can be manufactured with the available components in stock. In a pre-experiment, a flat solution structure appeared to be adequate for this problem. The two isomorphic problems are classification problems: the bibliographical references problem and the postage problem. The bibliographical references problem prints bibliographical references, e.g. articles from reviews or books, in accordance with current standards. The postage problem calculates the postage of various postal objects, e.g. parcels or letters, where the postage rate changes according to the object's type and the sending mode. In a pre-experiment, a hierarchical solution structure appeared to be adequate for these two problems.

---

[2] Data collected in a categorisation task allowed us to verify that these problems were isomorphic.

*Procedure*

The experts were asked to perform a design task. They used their familiar programming environment and their usual Smalltalk version (either 4.1 or 2.5). They were asked to think aloud while performing their task. We recorded[3] the order in which the elements of the solutions were generated. Each subject had to solve the three problems. The order of problems presentation was determined by random sampling without replacement among the 6 possible orders. No time limit was given for the task. It generally lasted from 4 to 5 hours.

*Method of analysis*

The successive drafts of solutions produced by the subjects, in particular, the order of generating methods were analysed. We focused on which kinds of relationship linked methods generated consecutively. The following links were distinguished:

— FUNCTIONAL SIMILARITY: the subject successively generates the code of methods which are functionally similar (for example: the methods of initialisation). Within this kind of relationship, we distinguished whether methods defined in a row belonged to the same class (within classes) or belonged to different classes (between classes). We have the following two kinds of links:
  ◊ functional similarity within classes (FSIntra): for example, the subject generates the code of an initialisation method in ClassA and another initialisation method in the same class.
  ◊ functional similarity between classes (FSInter): for example, the subject generates the code of an initialisation method in ClassA and another initialisation method in ClassB.

— MESSAGE PASSING RELATIONSHIP: the subject successively produces the code of methods which are related to each other in the message passing graph. We distinguished whether methods defined in a row belonged to the same class (within classes) or belonged to different classes (between classes). Thus we have the following two kinds of links:
  ◊ message passing within classes (MPIntra): for example, the subject generates the code of a MethodA in ClassA and the code of MethodB, which is called by MethodA, in the same class.
  ◊ message passing between classes (MPInter): for example, the subject generates the code of a MethodA in ClassA and the code of MethodB, which is called by MethodA, in ClassB.

---

[3] Under the Smalltalk environment a file in which changes made to the Smalltalk class library are recorded is created automatically.

- OTHER: the subject successively produces the code of methods which are not functionally similar and not in message passing relationship. We distinguished whether methods defined in a row belonged to the same class (within classes) or belonged to different classes (between classes). Thus we have the following two kinds of links:
    ◊ other within classes (OIntra): for example, the subject generates the code of a MethodA in ClassA and the code of MethodB in the same class. In this case, the subject completes the set or a subset of methods associated to a class.
    ◊ other between classes (OInter): for example, the subject generates the code of a MethodA in ClassA and the code of MethodB in ClassB. In this case the subject has already completed the set or a subset of methods associated to a class and moves on another class.

Links between successively defined methods have been coded on the basis of these categories. In this coding method, we did not take into account the deletion of methods and the syntactic modification of methods already created. One subject had a tool which generated some methods automatically. We did not take these methods into account in coding the links. The programs produced by the subjects have been represented according to two views: the message passing graph and the hierarchical structure of classes. This helped us to code the order in which lines of code were generated and to categorise the links between consecutively generated methods.

**Results**

*Program structures*

Table 1, Table 2, and Table 3 present quantitative data on the solutions produced by subjects for each problem. Program structures differed according to the problem type. As expected, most programs produced for the management problem had a flat structure and most programs produced for the two isomorphic problems had a hierarchical structure.

«INSERT TABLE 1 ABOUT HERE»

«INSERT TABLE 2 ABOUT HERE»

«INSERT TABLE 3 ABOUT HERE»

For the management problem, the decomposition of domain entities into classes did not require a class system using an inheritance hierarchy since the domain entities belong to different natural categories. The class structure is flat with horizontal communication between classes. For this problem, all the programs produced, except for the atypical solution produced by E4, had a flat structure. In these solutions, the communication between objects is mostly horizontal. This means that message calls are made mostly between sibling classes. Figure 1 shows an example of a flat solution for this problem.

« INSERT FIGURE 1 ABOUT HERE »

In contrast, for the two isomorphic problems, the domain entities belong to the same natural category. In consequence, the entities (or the object classes) are organised hierarchically. This hierarchical organization influences the way objects communicate. Communication between classes is vertical, mainly between superclasses and subclasses. All the programs produced, except the atypical solution produced by E2 for the postage problem, had a hierarchical structure. In these solutions, the communication between objects is mostly vertical. This means that message calls are made mostly between classes and their superclasses. Figure 2 shows an example of a hierarchical solution for the bibliographical references problem.

« INSERT FIGURE 2 ABOUT HERE »

To sum up, two program structures may be distinguished : a flat program structure with horizontal communication between classes for the management problem versus a hierarchical program structure with vertical communication between classes for the two isomorphic problems.

*Organization of the design activity*

The organization of the design activity consisted in identifying the classes first, then identifying the methods. Before starting coding in the environment, the subjects generally design the conceptual model of classes on paper. Under the environment, they developed the description of one or several classes (i.e. class name, superclass name, attribute names) before coding its or their methods. This order of code generation is constrained by the Smalltalk environment. It is not possible to create a method outside of an existing class thus the class description has to be created first. We observed that, when a method was generated, the subjects coded its name and the body of this method at the same time. They did not create the name, then generate the body of the method later on (after other activities).

« INSERT TABLE 4 ABOUT HERE »

« INSERT TABLE 5 ABOUT HERE »

« INSERT TABLE 6 ABOUT HERE »

Table 4, Table 5, and Table 6 present the number of alterations of class descriptions and methods made by the subjects for each problem. Corrections refer

to syntactic modifications. We observed that very few methods were deleted or moved from one class to another. This is consistent with results found in our previous study (Détienne 1995). We observed that experts revised less often the structure of their solutions than novices, in particular the association between methods and classes.

We observed episodes of code reuse of the style 'copy/edit'. The use of this style of code reuse, though not encouraged by the proponents of OOP, has already been well documented in the literature (Détienne 1991; Lange and Moher 1989; Rosson and Carroll 1993). In our study, we observed several such episodes when the subjects developed in a row several methods performing the same functionality and associated to the same class. These episodes were coded by the links 'functional similarity intra'. However, most of the time, the subjects preferred to write directly the code of the new method specially when the length of the method (and thus the cost of writing it) compared to copying a source/editing it, was low.

Copy/edit episodes did not occur when the source method and the target methods were associated to different classes. This is probably due to the cost of browsing the program in as much as our subjects used only one browser[4]. With one browser, even if a subject already knows what and where the source is, he/she must use the browser for doing the following actions: select the source class, select the category of method in which the source method is, select the source method. Then, he/she is able to copy the source code. He/she has to use the browser again for: selecting the target class, selecting the category of method in which the target method should be. Then the subject is able to create a target method by pasting the source method. The fact that we did not observe such copy/edit episodes is probably due to the cost of doing such a sequence of actions with one browser.

*Transitions between and within classes*

The order of generating methods, and, in particular, the relationship between methods generated consecutively, give information about the plan which guides the design activity.

« INSERT TABLE 7 ABOUT HERE »

Table 7 presents, for each problem, the number and percentages of transitions between classes according to the relationships between methods generated consecutively: either functional similarity or message passing, or other.

For the two isomorphic problems, the transitions between classes are mostly made according to the 'functional similarity' and to the 'other' category of links. For the management problem, the transitions between classes are mostly made according to the functional similarity and the message passing structure.

As two subjects' solutions were atypical, we recalculated the percentages without these protocols (numbers in brackets in table 7). For the postage problem, the majority of transitions between classes are made according to the functional similarity whereas for the management problem, the majority of transitions between classes are made according to the message passing structure. These results suggest that, for the postage problem, the generation of methods is centred around the functions, whereas, for the management problem, the generation of methods is guided by the procedure. In this case, methods are generated according to the message passing structure.

---

[4] This was not a constraint given by the experimentator.

« INSERT TABLE 8 ABOUT HERE »

Table 8 presents, for each problem, the number and percentage of transitions within classes according to the relationships between methods generated consecutively: either functional similarity or message passing or other.

For the bibliographical references problem, the transitions within classes are mostly made according to the 'functional similarity' and to the 'other' category of links. This means that, within a class, the subjects tend to consecutively develop methods which perform the same function, for example all the 'initialization' methods, then another function, for example all the 'reading access' methods, and so on.

For the postage problem, the transitions within classes are mostly made according to the functional similarity. This trend is even greater if we do not take into account the protocol with the atypical solution. This means that the subjects tend to consecutively develop methods which perform the same function, for example all the 'initialization' methods associated to a class.

For the management problem, the transitions within classes are mostly made according to the functional similarity and the message passing structure. However, without taking into account the protocol with the atypical solution, the majority of transitions within classes are made according to the message passing structure. This means that subjects tend to consecutively develop the methods according to their calling links in one class.

*Design strategies*

From our analysis of transitions between and within classes, we constructed a model of various design strategies which may be involved in the development of OO programs. Three strategies were distinguished: the function-centred strategy, the procedure-centred strategy, and the object-centred strategy. According to the function-centred strategy, the functions are prominent in the representation guiding the design activity, and objects are subordinate to functions. According to the procedure-centred strategy, the representation of the procedure guides the design activity. According to the object-centred strategy, the objects are prominent in the representation guiding the design activity, and functions as well as procedures are subordinate to objects.

*Function-centred strategy*

According to the function-centred strategy, the functions are prominent in the representation guiding the design activity, and objects are subordinate to functions. The subjects tend to follow plans in which *functions are central*. They develop one function for several objects, then another function for several objects and so on. Examples of functions are reading access, writing access, accessing, initialization, printing, etc. The following excerpt, in Table 9, illustrates this strategy. This strategy is reflected by a majority of links of the type 'functional similarity intra' (FSINTRA) and 'functional similarity inter' (FSINTER) as shown in the left column of Table 9.

« INSERT TABLE 9 ABOUT HERE »

In this example, the subject develops the methods which perform the function 'accessing' for the JOURNALPOSTAGE class, then the methods which perform this function for the PARCELPOSTAGE class, then the methods which perform this function for the ADMINISTRATIVEPOSTAGE class. These methods, performing the function 'accessing', return constant numbers which characterise each postal object and which are important in calculating their postage rates.

The function-centred strategy is reflected in the subjects' protocols by a majority of transitions between and within classes according to the functional similarity. This strategy was observed for the postage problem. For this problem, the transitions, whether between and within classes, are mostly made according to the functional similarity as shown in table 7 and table 8. We verified that this general trend reflects the strategies of the subjects taken individually. The analysis of individual protocols shows that E1, E3 and E4 have more than 50% (between 53% and 83%) of their links in the FS category. Their strategy is function-centred.

The individual protocol of E2 shows that this subject used a mixed strategy which followed the three types of links (functions, message passing and other). It should be remembered that this subject was the one who developed the atypical solution (the flat one compared to the hierarchical solution developed by the three other subjects).

To summarize: our result suggests that the majority of subjects tend to follow plans in which *functions are central* for developing the solution to the postage problem. It should be noted that we could assume that the prominence of this strategy for the postage problem is simply due to a great number of attributes (i.e. instance variables) which the methods of 'reading access' and 'writing access' are associated to. This would explain the use of functional similarity links for developing the 'reading access' and 'writing access' methods. This interpretation is not supported by our data. In fact this strategy is mostly used for the postage problem which has less attributes than the bibliographical references problem and the management problem: a total of 19 attributes are developed for this problem as compared to 54 and 23 respectively for the two other problems (as shown in Table 1, Table 2, and Table 3).

*Procedure-centred strategy*

According to the procedure-centred strategy, the representation of the procedure guides the design activity. The subjects tend to follow plans in which *methods implied in a procedure are organised according to their calling structure*. The structure of the procedure guides the program development. Actions of this procedure are attached to objects. The following excerpt, in Table 10, illustrates this strategy.

« INSERT TABLE 10 ABOUT HERE »

In this example, the subject develops in a row several methods (HowManyPossible in the CAR Class, howManyPossibleFor:aCar in the COMPONENT Class and cars.quantityTowards: aCar in the COMPONENT Class) implied in a procedure for calculating the number of cars which can be manufactured with the available components in stock. The method developed first calls the method developed secondly which calls the method developed in the third place. These transitions are coded by the links 'message passing intra' (MPINTRA) and 'message passing inter' (MPINTER) as shown in the left column of Table 10.

The procedure-centred strategy is reflected by a majority of transitions between and within classes according to the message passing structure. This strategy was observed for the management problem. For this problem, the transitions, whether between and within classes, are mostly made according to the message passing structure. We verified that this general trend reflects the strategies of the subjects taken individually. The analysis of individual protocols show that E1, E2 and E3 have more than 67% (between 67% and 73%) of their links in the MP category. Their strategy is procedure-centred.

The individual protocol of E4 shows that this subject followed a function-centred strategy: 61% of the links are in the FS category. It should be reminded that this subject was the one who developed an atypical solution: the hierarchical one compared to the flat solutions developed by the three other subjects. The strategy of this subject is the same as the one used by the three subjects who developed a hierarchical solution for solving the postage problem, as developed in the last section.

To summarize: our result suggests that a majority of subjects tend to follow plans in which *methods implied in a procedure are organised according to their calling structure* for the management problem.

*Object-centred strategy*

According to the object-centred strategy, the objects are prominent in the representation guiding the design activity, and functions as well as procedures are subordinate to objects.

This strategy was involved in solving one of the isomorphic problems, the bibliographical references problem. For this problem, we observed that the transitions, whether between and within classes, are made according to three kinds of links : functional similarity, message passing structure, and other. Taking the subjects individually, it was shown that three subjects (E1, E3 and E4) used a majority of O links (from 47% to 54%) and the fourth subject (E2) used a majority of FS links (45%) followed by O links (34%). These data were difficult to interpret without analysing the individual protocols in greater depth. This analysis suggested that the subjects tend to follow a plan in which several typical functions or in which several typical calling links are associated to a generic object. This means that *the objects are central*.

We are going first to illustrate the use of a plan in which several typical functions are associated to a generic object. For example, functions such as initialisation, reading access, writing access, and creation are associated to an object. Shifting from one class to another is triggered when several functions have been implemented in the previous class; then the subject tends to implement the same functions in another class. The objects are prominent in the representation guiding the design activity and functions are subordinate to objects. The following excerpt, in Table 11, illustrates this strategy.

« INSERT TABLE 11 ABOUT HERE »

In this example, the subject develops several methods which perform the initialisation function and a method which perform the printing function in the DOCUMENT Class, then develops methods which perform these functions in the ARTICLE Class, then again in the REPORT Class. The transitions are of the type 'functional similarity intra' (FSINTRA) and 'other intra and inter' (OINTRA and OINTER). It should be noted that the fact that a sequence of typical functions is repeated from one class to another is not directly highlighted by our coding system in as much as the links are coded as a binary relation (from one method to the next generated method). It is a limitation of our coding system.

In addition to typical functions associated to a generic object, typical calling links between methods may be associated to the object. We observed that transitions within classes can be made according to this calling link, i.e. message passing link, as well as according to the functional similarity between methods. In this case, the subjects develop several methods by following their calling structure in ClassA and, later on, develop several similar methods by following their calling structure again in ClassB, and so on. This suggests that the subjects follow a plan in which *several typical functions as well as typical calling structures are attached to a generic object*. The following excerpt, in Table 12, illustrates this latter aspect of the object-centred strategy.

« INSERT TABLE 12 ABOUT HERE »

In this example, the subject develops the methods performing the printing function and the reading access function for the BOOK Class, then methods performing these functions for a sibling class, the REPORT Class. There are no calling links between the methods of the BOOK Class and methods of the REPORT class. But there is a calling structure between printing and reading methods of the BOOK class and the same calling structure between the printing and reading access methods of the REPORT class.

Again, it should be noted that the fact that a sequence of calling links is repeated from one class to another is not directly highlighted by our coding system in as much as the links are coded as a binary relation.

**Discussion**

Three strategies were observed in this study: a function-centred strategy, a procedure-centred strategy, and an object-centred strategy. According to the function-centred strategy, the functions are prominent in the representation guiding the design activity, and objects are subordinate to functions. According to the procedure-centred strategy, the representation of the procedure guides the design activity. According to the object-centred strategy, the objects are prominent in the representation guiding the design activity, and functions as well as procedures are subordinate to objects. Similar search strategies have been described in the study conducted by Rist (to appear): a design strategy based on roles, a design strategy based on goals and a design strategy based on objects. A design strategy based on roles expands one role at the time. Similar roles exist across goals and/or objects. In this paper, this strategy refers to the 'functional similarity' category of links. A design strategy based on goals expands one goal at the time, varying the role and the object. In this paper, this strategy refers to the 'message passing' category of links. A design strategy based on objects expands one object at the time, varying the roles and goals relevant to those objects. In this paper, this strategy refers to the 'other' category of links.

*Program structures and design strategies*

A question was to analyse whether or not the use of design strategies is related to different solution structures. For the management problem, the solutions produced by subjects, except one (E4), had a flat structure. In this type of solution, there is only one level below the Object class, and all the classes are defined at this level. The natures of these classes are quite different from each other and communication between classes is horizontal. For the two isomorphic problems, the subjects, except one (E2) for one problem, developed solutions with hierarchical structures. In these solutions, there are 2 or 3 levels in the hierarchy of classes. There is no communication between leaves of the tree. These sub-classes communicate only vertically with their superclasses.

We could interpret our results in relation to the solution structures developed by experts. We observed that, when the structure of the solution is flat with horizontal communication, as for the management problem, the plan which guides the design activity would tend to be procedural. It is composed of methods implied in a procedure, with these methods organised according to their calling links. It is assumed that an initial representation of the solution with a flat structure and horizontal communication between classes would trigger the procedure-centred strategy.

We observed that, when the structure of the solution is hierarchical with vertical communication, as for the two isomorphic problems, the plan which guides the design activity would tend to be declarative. The strategy is either object-centred, i.e. objects are prevalent and functions are attached to objects, or function-centred, i.e. functions are prevalent and objects are attached to functions. It is assumed that an initial representation of the solution with a hierarchical structure and vertical communication between classes would trigger the function-centred strategy or the object-centred strategy.

To conclude, whatever the design paradigm, two main plans may guide the design activity. Either the plan is declarative: static characteristics like objects, main goals, and typical functions (roles) guide the solution development. Or the plan is procedural: dynamic characteristics of the procedure guide the solution development.

In procedural design, the study of Hoc (1983) showed that the use of one type of plan could be linked to the type of problem at hand. For procedural problems, the program structure is strongly constrained by the procedure structure; this structure guides the solution development and the plan is procedural. For declarative problems, the program structure is strongly constrained by the data structure; this structure guides the solution development and the plan is declarative. In OOD, the use of one plan rather than another does not seem to be determined by the same dimension of the problem (Détienne 1995). A dimension more relevant to OOD may characterize not only the data structure (or objects), on one side, or the procedure, on the other side, but rather the way they are associated. In the present study, our results tend to show that the plan was declarative for developing hierarchical solution structures with vertical communication between objects and procedural for developing flat solution structures with horizontal communication between objects. This suggests that a dimension of problems more relevant for OOD is the organization of objects and the way objects communicate within this structure: either vertically in a hierarchical organization or horizontally in a flat organization.

Hierarchical solution structures are typical in OO design. For developing a solution with a hierarchical structure and vertical communication between classes, two design strategies have been used by our experts, either the object-centred strategy or the function-centred strategy. The use of a declarative plan, either function-centred or object-centred, highlights the role of main goals and objects of the problem as well as typical functions (roles) in the problem decomposition. The important role of objects in this decomposition is consistent with the claims made by advocates of OOD. Advocates of OO design have made strong claims about the naturalness, ease of use, and power of this design approach (Meyer 1988; Rosson and Alpert 1990). The theoretical argument in support of OO design is that objects are clear and visible entities in the problem domain, are represented as explicit or 'first-class' entities in the solution domain, and thus the mapping between the problem and solution domains is simple and clear. Flat solution structures are more typical of procedural design. The use of a procedural plan highlights the role of the dynamic structure of the procedure in the problem decomposition. In the present study, our results tend to show that the OOD experts use a procedure-centred strategy for developing a flat solution structure with horizontal communication between classes.

For any large software development, it is likely that the solution has parts with hierarchical structure and vertical communication and parts with flat structure and horizontal communication. In this case, we could assume that various design strategies, either object or function or procedure-centred, are used for developing parts which have different structures.

It is worth noting that only the object-centred strategy was observed in a previous study (Détienne 1995) with experts in OOD. A difference between these studies is the OOP language and device used, $CO_2$ in the previous study and Smalltalk in the present study. It is likely that some Smalltalk environment characteristics may have triggered the use of the function-centred strategy. Under this environment, methods performing the same function, e.g. initialisation, can be grouped by the programmer into a category of methods with the label 'init', for example. According to the function-centred strategy, the subjects tend to develop one category of methods for one object, then for other objects, then move on another category of methods and so on.

Another difference between these studies is that in the $CO_2$ study, the experts tend to develop the structure of the solution (names and structures of classes, names of methods) before developing the code of methods. In the present study the experts developed the names and the code of methods at the same time: the coding of methods was not postponed. This may explain the difference of results. In the previous study, although we observed mainly the use of the object-centred strategy in the preliminary planning phase, we showed that the functional similarity may be very important in the coding phase as we observed many code reuse episodes based on similar functionality.

*Perspectives of research*

The representations constructed in a design task provide insight into knowledge structures possessed by experts in OOD. According to schema-based models of programming, experienced programmers are assumed to possess in memory schemas (Détienne 1990b, c; Rist 1986; Robertson and Yu 1990; Soloway and Ehrlich 1984) which are abstract knowledge structures they have constructed through practice in their domain of expertise.

In the procedural programming domain, a schema may be described as a set of actions, i.e., execution steps, with some constraints on the order in which actions are executed. Actions can be categorised as typical roles. Rist distinguishes between various roles like input, calculate, output. One or several lines of a program may be represented as part of one role and this role may be represented as part of one programming plan. A programming plan represents portions of code which achieve a common goal and a program is viewed as a set of complex and basic plans which are merged together to achieve the problem goal (Rist 1991). When a programming plan is memorised as a knowledge structure, it is called a plan schema (Rist 1986) or plan (Soloway et al. 1982) or schema (Détienne 1990a).

Whereas evidence supporting the hypothesis that experts possess schemas has been found in various studies on procedural design, the nature of knowledge constructed through practice with an object-oriented programming language remains unclear. The analysis of design strategies provides insight into the knowledge used in a design activity. Design strategies accounting for the order in which parts of programs are generated by the subjects may reveal the use of schemas. If a schema is evoked, elements of this schema would be made available in their schema order. Rist (1991) showed that evidence for schema retrieval consisted of actions appearing in their executable or schema order.

In a general way, we will assume that knowledge structures possessed by experts in OOD would integrate objects and procedures which use these objects to achieve goals. According to the prominence of the procedure's characteristics or the object's characteristics, two alternative views of schemas are possible. One view is to assume that a schema is centred around one main procedure to achieve a goal. In this case, it would integrate characteristics of the procedure, and characteristics of objects used by this procedure. In our design task, the procedure-centred strategy suggests the use of such knowledge structure. Another view is to assume that a schema is centred around one main object. In this case, it would integrate the characteristics of this object and, in addition, actions linked to this object. This schema would integrate actions related to several procedures but linked to the same object. Actions may represent typical functions or roles which are independent of the problem domain. In our design task, the object-centred strategy and the function-centred strategy suggest the use of such knowledge structures. The object-centred strategy would consist in instantiating completely such a schema for one object, then for another object of the solution and so on. The function-centred strategy would consist in instantiating partially such a schema for one object, for example, for the 'initialization' methods, then partially instantiating the schema for another object (and for the same function) and so on.

More empirical studies could be conducted in this perspective of research. Data of our study provided us with only weak support to our hypotheses on knowledge organisation. Knowledge elicitation tasks could provide us with data to evaluate these hypotheses. We are currently following this direction of research.

*Limitations and implications of this study*

To conclude, we should point out limitations of this study. Even though this study provides interesting insight into the design strategies and knowledge used in the design activity with an object-oriented language, it is necessary to conduct further studies with a larger number of subjects so as to generalise our results. Our analysis has also highlighted some limitations of our coding system; it could be extended to take into account n-ary relations and different types of units, e.g. methods and groupings of methods.

An issue is to analyse further which characteristics of the design situation may trigger the use of one strategy rather than another. This issue is particularly important for developing support to the design activity which would take into account which conditions, external and internal to the subjects, trigger a particular strategy. For problems more complex than those used in this study, it is likely that the experts may use a combination of various strategies. For example, the object-centred strategy and the function-centred strategy could be used locally when reuse by inheritance is necessary. As observed in a previous study (Détienne 1991), the super class, used as a source in the reuse activity, could be developed first according to an object-centred strategy, then the subclasses, acting as the targets, could be developed with a function centred-strategy. This last strategy would allow the subject to specialise a function in all the subclasses, then to specialise another function in all the subclasses again. In our previous study, we observed that this behaviour, referred to as 'reuse in a row', avoided errors like omissions of changes.

These results have implications for HCI systems, in particular for systems assisting experts in their design activity. The results of this study highlight the diversity of strategies used by programmers developing programs with an OOP language. Support systems should take these various strategies into account. It is important to provide visualisation of programs which support these various strategies: a representation of the message passing graph for the procedure-centred strategy, and a representation of the organisation of classes and of roles for the object-centred strategy and the function-centred strategy. The environment could also be sensitive to certain conditions triggering one strategy or another. For example, if the subject develops a flat conceptual structure, then a representation of the message passing graph could be the visualisation proposed by default in as much as it supports the procedure-centred strategy which may be triggered under these conditions.

More specifically, our results suggest the following improvements of the Smalltalk environment[5]. As regards the visualisation aspect, on one hand, the visualisation of roles is already possible in Smalltalk by the concept of category of methods. On the other hand, there is no visualisation of the message calling graph. There is only a means to visualise where a particular method is called or implemented. We have observed in this study that methods with a particular calling structure are repeated in a different class. It could be made possible to reuse the methods implied in a particular calling structure if the programmer could select this structure on the basis of a message calling graph. Providing the programmer with a means to visualise the message calling graph would support the design activity. The inheritance graph exhibits only classes. There is no global visualisation of classes and methods associated to classes. It is only possible to visualise the methods associated to one particular class by selecting this class in the browser and « class refs' from the menu of the class names pane. It should be noted that some research for providing the designer with the static structure as well as the dynamic structure of Smalltalk programs has been made (Böcker and Herczeg 1990).

*Acknowledgements*

We would like to express our thanks to Isabelle Borne for her collaboration in this study, and to Thomas Green, Robert Rist and an anonymous reviewer for their helpful comments on a previous draft of this paper.

---

[5] These remarks are based on the version of the Smalltalk environment we know as the last one. It may be possible that more recent environments have characteristics we are not conscious of.

## References

Adelson, B. and E. Soloway, 1985. The role of domain experience in software design. IEEE Transactions on Software Engineering 11, 1351-1360.

Böcker, H-D. and J. Herczeg, 1990. 'Browsing through program execution'. In: D. Diaper, D. Gilmore, G. Cockton and B. Shacker (eds.), Human Computer Interaction, proceedings of INTERACT'90, North Holland, 991-996.

Burkhardt, J-M. and F. Détienne, 1995. 'An emprical study of software reuse by experts in object-oriented design'. In: K. Nordby, P.H. Helmersen, D.J. Gilmore and S.A. Arnesen (eds.), Human Computer Interaction, proceedings of INTERACT'95. Chapman & Hall, 133-138.

Chatel, S., F. Détienne, and I. Borne, 1992. 'Transfer among programming languages: an assessment of various indicators'. In: F. Détienne (ed.), Proceedings of the Fifth Workshop of the Psychology of Programming Interest Group. INRIA, 261-272.

Davies, S. P., 1993. Models and theories of programming strategy. International Journal of Man-Machine Studies 39, 237-267.

Détienne, F., 1990a. 'Difficulties in Designing with an object-oriented language: an empirical study'. In: D. Diaper, D. Gilmore, G. Cockton and B. Shacker (eds.), Human Computer Interaction, proceedings of INTERACT'90. North Holland, 971-976.

Détienne, F., 1990b. 'Expert programming knowledge: a schema-based approach'. In: J.-M. Hoc, T.R.G. Green, R. Samurçay and D. Gilmore (eds.), Psychology of programming. London: Academic Press, People and Computer Series, 205-222.

Détienne, F., 1990c. 'Program understanding and knowledge organization: the influence of acquired schemas'. In: P. Falzon (ed.), Cognitive Ergonomics: Understanding, Learning and Designing Human-Computer Interaction. London: Academic Press, 245-256.

Détienne, F., 1991. 'Reasoning from a schema and from an analog in software code reuse'. In: J. Koenemann-Belliveau, T. Moher, and S.P. Robertson (eds.), Empirical studies of programmers, Fourth Workshop. Norwood, NJ: Ablex Publishing Corporation, 5-22.

Détienne, F., 1995. Design strategies and knowledge in object-oriented programming: effects of experience. HCI Journal 10, 129-170.

Guindon, R., 1990. Designing the design process: exploiting opportunistic thoughts. Human-Computer Interaction 5, 305-344.

Hoc, J.-M., 1983. Une méthode de classification préalable des problèmes d'un domaine pour l'analyse des stratégies de résolution: la programmation informatique chez des professionnels. Le Travail Humain 46, 205-217.

Kim, J. and J. Lerch, 1992. 'Towards a model of cognitive process in logical design: comparing object-oriented and traditionnal functional decomposition software methodologies'. In: P. Bauersfeld, J. Bennett and G. Lynch (eds.), Proceedings of CHI'92 Conference on Human Factors in Computing Systems. ACM Press, 489-498.

Lange, B.M. and T.G. Moher, 1989. 'Some strategies of reuse in an object-oriented programming environment'. In K. Bice and C. Lewis (eds.), Proceedings of CHI'89 Conference on Human Factors in Computing Systems. ACM Press, 69-73.

Lewis, J.A., S.M. Henry, D.G. Kafura and R.S. Schulman, 1991. 'An empirical study of the object-oriented paradigm and software reuse'. Proceedings of Object-Oriented Programming, Systems and Applications. ACM Press: NY, 184-196.

Meyer, B. , 1988. Object-Oriented Software Construction. Englewood Cliffs, NJ: Prentice Hall.

Pennington, N. and B. Grabowski, 1990. 'The tasks of programming'. In: J.-M. Hoc, T.R.G. Green, R. Samurçay and D. Gilmore (eds.), Psychology of programming. London: Academic Press, Computer and People Series, 45-62.

Pennington, N., A. Lee, and B. Rehder, 1995. Cognitive activities and levels of abstraction in procedural and object-oriented design. HCI Journal 10, 171-226.

Ratcliff, B. and J.A. Siddiqi, 1985. An Empirical investigation into problem decomposition strategies used in program design. International Journal of Man-Machine Studies 22, 77-90.

Rist, R., 1986. 'Plans in programming: definition, demonstration, and development'. In: E. Soloway and S. Iyengar (eds.), Empirical Studies of Programmers: First Workshop. Norwood, NJ: Ablex Publishing Corporation, 28-47.

Rist, R., 1991. Knowledge creation and retrieval in program design: a comparison of novice and intermediate student programmers. Human-Computer Interaction 6, 1-46.

Rist, R. (to appear). 'System structure and design'. Empirical Studies of Programmers: SixthWorkshop, Washington DC, US, 5-7 January 1996.

Robertson, S.P. and C.C. Yu, 1990. Common cognitive representations of program code across tasks and languages. International Journal of Man-Machine Studies 33, 343-360.

Rosson, M.B. and S.R. Alpert, 1990. The cognitive consequences of object-oriented design. Human-Computer Interaction 5, 345-379.

Rosson, M.B. and J.M. Carroll, 1993. 'Active programming strategies in reuse'. Proceedings of ECOOP'93, Object-Oriented Programming. Berlin: Springer-Verlag. 4-18.

Soloway, E. and K. Ehrlich, 1984. Empirical studies of programming knowledge. IEEE Transactions on Software Engineering SE-10, 595-609.

Soloway, E., K. Ehrlich, and J. Bonar, 1982. 'Tapping into tacit programming knowledge'. Proceedings of CHI'82 Conference on Human Factors in Computing Systems. ACM Press, 52-57.

Visser, W. and J-M. Hoc, 1990. 'Expert software design strategies'. In: J.-M. Hoc, T.R.G. Green, R. Samurçay and D. Gilmore (eds.), Psychology of programming. London: Academic Press, Computer and People Series, 235-249.

# Tables and figures

Table 1
Characteristics of programs produced for the Management problem

|  | Experienced programmers in OOP | | | | |
| --- | --- | --- | --- | --- | --- |
| Characteristics | E1 | E2 | E3 | E4 | Total |
| Number of defined classes | 2 | 2 | 4 | 2 | 10 |
| Number of attributes (instance variables) | 0 | 10 | 10 | 3 | 23 |
| Hierarchical structure | no | no | no | yes | N/Y |
| Maximum depth of class hierarchization[d] | 1 | 1 | 1 | 2 | [1-2] |
| Number of defined methods | 4 | 9 | 4 (18) | 14 | 31 |

[d]Depth equal to 1 corresponds to flat structure
()Number of methods generated automatically

Table 2
Characteristics of programs produced for the Bibliographical references problem

| Characteristics | Experienced programmers in OOP | | | | Total |
| --- | --- | --- | --- | --- | --- |
| | E1 | E2 | E3 | E4 | |
| Number of defined classes | 6 | 6 | 12 | 6 | 30 |
| Number of attributes (instance variables) | 15 | 11 | 17 | 11 | 54 |
| Hierarchical structure | yes | yes | yes | yes | Y |
| Maximum depth of class hierarchization[d] | 2 | 3 | 3 | 3 | [2-3] |
| Number of defined methods | 17 | 41 | 27 (44) | 17 | 102 |

[d]Depth equal to 1 corresponds to flat structure
()Number of methods generated automatically

Table 3
Characteristics of programs produced for the Postage problem

|  | Experienced programmers in OOP | | | | |
| --- | --- | --- | --- | --- | --- |
| Characteristics | E1 | E2 | E3 | E4 | Total |
| Number of defined classes | 5 | 2 | 12 | 5 | 24 |
| Number of attributes (instance variables) | 4 | 5 | 8 | 2 | 19 |
| Hierarchical structure | yes | no | yes | yes | Y/N |
| Maximum depth of class hierarchization[d] | 3 | 1 | 3 | 2 | [1-3] |
| Number of defined methods | 19 | 24 | 16 (37) | 24 | 83 |

[d]Depth equal to 1 corresponds to flat structure
()Number of methods generated automatically

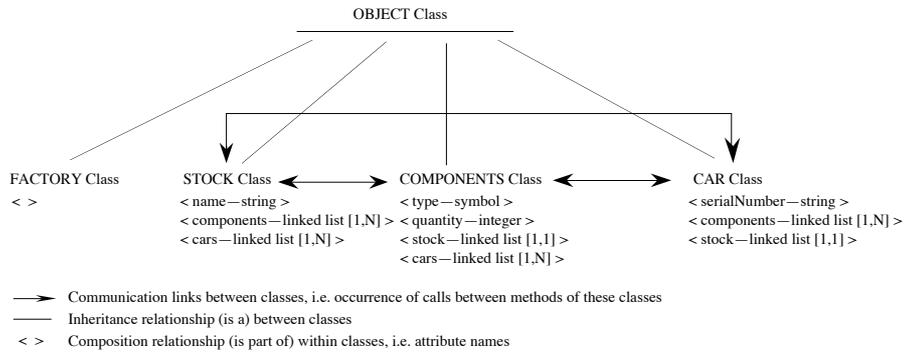

Fig. 1. Excerpt of flat structure with horizontal communication between classes for the Management problem

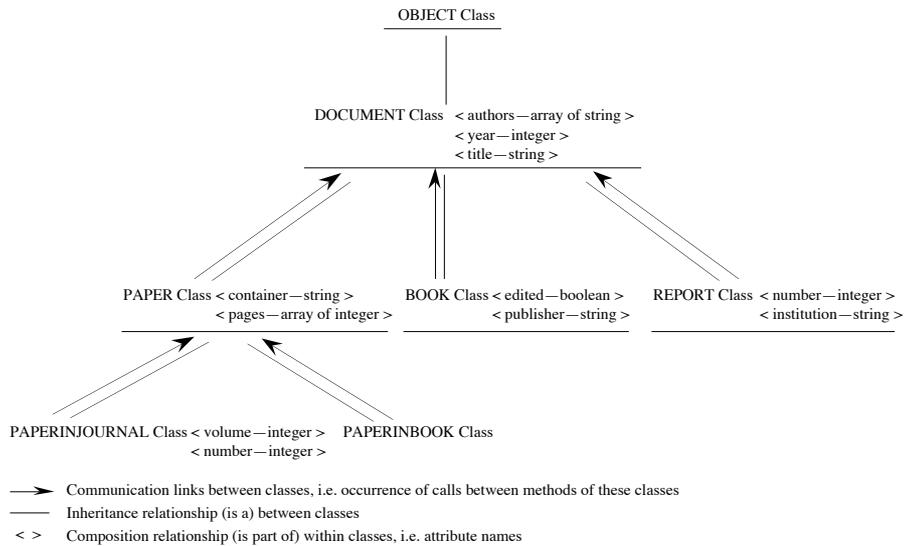

Fig. 2. Excerpt of hierarchical structure with vertical communication between classes for the Bibliographical references problem

Table 4
Number of alterations made for the Management problem

| Type of alterations | | Experienced programmers in OOP | | | | |
| --- | --- | --- | --- | --- | --- | --- |
| | | E1 | E2 | E3 | E4 | Total |
| Class descriptions | modified | 0 | 0 | 0 | 1 | 1 |
| | suppressed | 0 | 0 | 0 | 0 | 0 |
| | corrected | 0 | 0 | 0 | 0 | 0 |
| Total number of altered class descriptions | | 0 | 0 | 0 | 1 | 1 |
| Methods | modified | 0 | 3 | 2 | 4 | 9 |
| | suppressed | 0 | 0 | 0 | 2 | 2 |
| | corrected | 3 | 4 | 0 | 4 | 11 |
| Total number of altered methods | | 3 | 7 | 2 | 10 | 22 |

Table 5
Number of alterations made for the Bibliographical references problem

|  |  | Experienced programmers in OOP | | | | |
|---|---|---|---|---|---|---|
| Type of alterations |  | E1 | E2 | E3 | E4 | Total |
| Class descriptions | modified | 0 | 1 | 4 | 1 | 6 |
|  | suppressed | 0 | 0 | 0 | 0 | 0 |
|  | corrected | 3 | 0 | 0 | 0 | 3 |
| Total number of altered class descriptions |  | 3 | 1 | 4 | 1 | 9 |
| Methods | modified | 0 | 3 | 4 | 2 | 9 |
|  | suppressed | 1 | 1 | 0 | 1 | 3 |
|  | corrected | 2 | 9 | 0 | 3 | 14 |
| Total number of altered methods |  | 3 | 13 | 4 | 6 | 26 |

Table 6
Number of alterations made for the Postage problem

|  |  | Experienced programmers in OOP | | | | |
| --- | --- | --- | --- | --- | --- | --- |
| Type of alterations |  | E1 | E2 | E3 | E4 | Total |
| Class descriptions | modified | 0 | 2 | 5 | 0 | 7 |
|  | suppressed | 0 | 0 | 0 | 0 | 0 |
|  | corrected | 0 | 0 | 0 | 0 | 0 |
| Total number of altered class descriptions |  | 0 | 2 | 5 | 0 | 7 |
| Methods | modified | 0 | 1 | 2 | 0 | 3 |
|  | suppressed | 0 | 1 | 6 | 0 | 7 |
|  | corrected | 7 | 2 | 3 | 1 | 13 |
| Total number of altered methods |  | 7 | 4 | 11 | 1 | 23 |

Table 7
Number (#) and percentage (%) of transitions between classes according to the type of link per problem

| | ISOMORPHIC PROBLEMS | | | | | |
| --- | --- | --- | --- | --- | --- | --- |
| | BIBLIOG REF | | POSTAGE | | MANAGEMENT | |
| TYPE OF TRANSITIONS | # | % | # | % | # | % |
| Functional similarity | 20 | 34.5 | 15 (15) | 53.6 (62.5) | 6 (2) | 42.9 (25.0) |
| Message passing | 12 | 20.7 | 2 (1) | 7.1 (4.2) | 6 (6) | 42.9 (75.0) |
| Other | 26 | 44.8 | 11 (8) | 39.3 (33.3) | 2 (0) | 14.2 (0.0) |
| Overall transitions | 58 | 100 | 28 (24) | 100 (100) | 14 (8) | 100 (100) |

( ) Numbers and percentages without atypical solutions

Table 8
Number (#) and percentage (%) of transitions within classes according to the type of link per problem

| | ISOMORPHIC PROBLEMS | | | | | | | |
|---|---|---|---|---|---|---|---|---|
| | BIBLIOG REF | | POSTAGE | | | MANAGEMENT | | |
| TYPE OF TRANSITIONS | # | % | # | | % | # | | % |
| Functional similarity | 17 | 32.7 | 34 | (24) | 61.8 (70.6) | 7 | (0) | 31.8 (0.0) |
| Message passing | 13 | 25.0 | 11 | (6) | 20.0 (17.6) | 8 | (6) | 36.4 (66.7) |
| Other | 22 | 42.3 | 10 | (4) | 18.2 (11.8) | 7 | (3) | 31.8 (33.3) |
| Overall transitions | 52 | 100 | 55 | (34) | 100 (100) | 22 | (9) | 100 (100) |

( ) Numbers and percentages without atypical solutions

Table 9
Example of development centred around functions for the Postage problem

| LINKS | GENERATION ORDER | PROGRAM CODE | PLAN STRUCTURE |
|---|---|---|---|
| | [...] | ... | |
| | .12. | JOURNALPOSTAGE Class <> | OBJECT |
| FSINTRA | [13] | byAir | |
| FSINTRA | [14] | byRegistered | ACCESSING |
| FSINTRA | [15] | baseRate | |
| FSINTRA | [16] | weightRate | |
| **FSINTER** | [17] | weightUnit | |
| | .18. | PARCELPOSTAGE Class <> | OBJECT |
| FSINTRA | [19] | byAir | |
| FSINTRA | [20] | byRegistered | |
| FSINTRA | [21] | baseRate | ACCESSING |
| FSINTRA | [22] | weightRate | |
| **FSINTER** | [23] | weightUnit | |
| | .24. | ADMINISTRATIVEPOSTAGE Class <> | OBJECT |
| FSINTRA | [25] | byAir | |
| FSINTRA | [26] | byRegistered | |
| FSINTRA | [27] | baseRate | ACCESSING |
| FSINTRA | [28] | weightRate | |
| | [29] | weightUnit | |
| | [...] | ... | |

The number in square brackets [] corresponds to the generation order in which methods are developed. The indentation corresponds to the hierarchical structure of the final solution (i.e. inheritance relationship 'is a'). The angle brackets <> correspond to the attributes (i.e. instance variables) of the classes (i.e. composition relationship 'is part of').

Table 10
Example of development centred around sending message structure for the Management problem

| LINKS | GENERATION ORDER | PROGRAM CODE | | PLAN STRUCTURE |
|---|---|---|---|---|
| | [...] | ... | | |
| | .13. | CAR Class <serialNumber, components, stock> | | OBJECT |
| **MPINTER** | [25] | howManyPossible | | | CALCULATION |
| | .4. | COMPONENT Class <type, quantity, stock, cars> | | OBJECT |
| MPINTRA | [26] | howManyPossibleFor: aCar | | | CALCULATION |
| | [27] | cars.quantityTowards: aCar | | | ACCESSING |
| | [...] | ... | | |

The number in square brackets [] corresponds to the generation order in which methods are developed. The indentation corresponds to the hierarchical structure of the final solution (i.e. inheritance relationship 'is a'). The angle brackets <> correspond to the attributes (i.e. instance variables) of the classes (i.e. composition relationship 'is part of').

Table 11
Example of development centred around objects for the Bibliographical references problem

| LINKS | GENERATION ORDER | PROGRAM CODE | PLAN STRUCTURE | |
|---|---|---|---|---|
|  | .1. | DOCUMENT Class <authors, year, title> | OBJECT | |
| FSINTRA | [2] | authors: anArrayOfString | | INIT |
| FSINTRA | [3] | year: aString | | |
| OINTRA | [4] | title: aString | | |
| **OINTER** | [5] | print | | PRINTING |
|  | .6. | PAPER Class <reviewName, volume, number, pages> | OBJECT | |
| FSINTRA | [7] | reviewName: aString | | INIT |
| FSINTRA | [8] | volume: aNumber | | |
| FSINTRA | [9] | number: aNumber | | |
| OINTRA | [10] | pages: aNumber to: aNumber | | |
| **OINTER** | [11] | print | | PRINTING |
|  | .12. | REPORT Class <reportNumber, institution> | OBJECT | |
| FSINTRA | [13] | reportNumber: aNumber | | INIT |
| OINTRA | [14] | institution: aString | | |
|  | [15] | print | | PRINTING |
|  | [...] | ... | | |

The number in square brackets [] corresponds to the generation order in which methods are developed. The indentation corresponds to the hierarchical structure of the final solution (i.e. inheritance relationship 'is a'). The angle brackets <> correspond to the attributes (i.e. instance variables) of the classes (i.e. composition relationship 'is part of').

Table 12
Example of development centred around objects for the Bibliographical references problem

| LINKS | GENERATION ORDER | PROGRAM CODE | PLAN STRUCTURE |
|---|---|---|---|
| | [...] | ... | |
| | .24. | BOOK Class <edited, publisher> | OBJECT |
| MPINTRA | [25] | printOn: aSTream | |
| MPINTRA | [26] | printNotEdited: aStream | PRINTING |
| MPINTRA | [27] | printPublisher: aStream | |
| MPINTRA | [28] | publisher | READING ACCESS |
| OINTER | [29] | printEdited: aStream | PRINTING |
| | .30. | REPORT Class <reportNumber, institution> | OBJECT |
| FSINTRA | [31] | reportNumber | READING ACCESS |
| OINTRA | [32] | institution | |
| MPINTRA | [33] | printOn: aStream | |
| MPINTRA | [34] | printReportNumber: aStream | PRINTING |
| | [35] | printInstitution: aStream | |
| | [...] | ... | |

The number in square brackets [] corresponds to the generation order in which methods are developed. The indentation corresponds to the hierarchical structure of the final solution (i.e. inheritance relationship 'is a'). The angle brackets <> correspond to the attributes (i.e. instance variables) of the classes (i.e. composition relationship 'is part of').